# ErMn$_6$Sn$_6$: A Promising Kagome Antiferromagnetic Candidate for Room-Temperature Nernst Effect-based thermoelectrics

*Olajumoke Oluwatobiloba Emmanuel, Shuvankar Gupta[*] and Xianglin Ke[*]*

Olajumoke Oluwatobiloba Emmanuel, Shuvankar Gupta and Xianglin Ke

Department of Physics and Astronomy, Michigan State University, East Lansing, Michigan 48824-2320, USA

E-mail: kexiangl@msu.edu,  guptas40@msu.edu
("Olajumoke Oluwatobiloba Emmanuel and Shuvankar Gupta contributed equally to this work.")

Keywords: (Kagome magnets, Nerst effects, Topological Hall, Topological Nernst, Thermoelectric)

**Abstract text.** The Nernst effect, the generation of a traverse electric voltage in the presence of longitudinal thermal gradient, has garnered significant attention in the realm of magnetic topological materials due to its superior potential for thermoelectric applications. In this work, we investigate electronic and thermoelectric transport properties of a Kagome magnet ErMn$_6$Sn$_6$, a compound showing an incommensurate antiferromagnetic phase followed by a ferrimagnetic phase transition upon cooling. We show that in the antiferromagnetic phase ErMn$_6$Sn$_6$ exhibits both topological Nernst effect and anomalous Nernst effect, analogous to the electric Hall effects, with the Nernst coefficient reaching 1.71 µV K$^{-1}$ at 300 K and 3 T. This value surpasses that of most of previously reported state-of-the-art canted antiferromagnetic materials and is comparable to recently reported other members of RMn$_6$Sn$_6$ (R = rare-earth, Y, Lu, Sc) compounds, which makes ErMn$_6$Sn$_6$ a promising candidate for advancing the development of Nernst effect-based thermoelectric devices.

## 1. Introduction

Magnets with a Kagome lattice present a rich platform for the investigation of various extraordinary electronic and magnetic phenomena arising from the Kagome lattice geometry. [1–4] In particular, recently metallic $RMn_6Sn_6$ (R = rare-earth, Y, Lu, Sc) magnets, in which manganese atoms form a Kagome lattice, have been characterized with distinctive electronic features such as Dirac points, van Hove singularities, flat bands and the resulting physical properties. [5–8] For instance, $TbMn_6Sn_6$ exhibits a fascinating combination of a Chern gap [9] and magneto-thermal properties [10,11], along with a skyrmion spin lattice [12], providing valuable insights into topological electronic properties and spin textures. And $YMn_6Sn_6$ exhibits a room-temperature topological Hall effect, showcasing its potential for spintronic applications. [13,14] More broadly, the $RMn_6Sn_6$ series were reported to show topological transport properties that are attributable to nontrivial Berry phase arising from the Kagome lattice, highlighting the roles of lattice geometry on the topological electronic properties of these compounds. [15]

In recent years, there has been growing interest in studying anomalous Nernst effect (ANE) in magnetic topological materials, partially driven by its distinct advantages over conventional Seebeck-based devices in thermoelectric applications. [10,16–21] The ANE, analogous to the anomalous Hall effect (AHE), manifests as a transverse thermoelectric voltage perpendicular to both the temperature gradient and the applied magnetic field, all of which are mutually orthogonal. Unlike the AHE which is determined by the Berry curvature of electronic bands over the whole Fermi-sea, the ANE is particularly sensitive to Berry curvature of electronic bands near the Fermi level, making it a potent indicator of thermoelectric efficiency in materials with significant Berry curvature, such as certain ferromagnetic (FM) compounds. [20–22] However, the presence of stray magnetic fields in FM materials poses stability challenges for ANE-based devices, prompting exploration into antiferromagnetic (AFM) materials like MnGe, [23] $Mn_3Sn$, [24] $Mn_3Ge$, [25] and $YbMnBi_2$, [26] which exhibit noncollinear magnetic structures and lower susceptibility to stray fields. Despite initial successes, most of AFM materials investigated have shown relatively low ANE values, with notable exceptions like YbMnBi$_2$ which displays a high ANE albeit restricted to low temperatures. [26] To advance practical applications, identifying AFM materials with high ANE at room temperature is crucial. Recent studies have reported a large ANE in AFM compounds YMn$_6$Sn$_6$ [27] and ScMn$_6$Sn$_6$ [28] as well as in a ferrimagnetic (FIM) compound TbMn$_6$Sn$_6$ [10,11], opening new directions for exploring ANE-based compounds within the RMn$_6$Sn$_6$ family. Notably, ErMn$_6$Sn$_6$ presents a unique case due to its competing Er-Mn and Mn-Mn couplings along the *c*-axis, [29] which give rise to FIM and incommensurate AFM phases in different temperature regimes. [30] Note that ErMn$_6$Sn$_6$ is on the border between (Y, Sc, Lu)Mn$_6$Sn$_6$ which have an incommensurate AFM ground state and (Ho, Dy, Tb, Td)Mn$_6$Sn$_6$ which are ferrimagnets, possessing a ferrimagnetic ground state preceded by an AFM phase with incommensurate spin structure of both Mn and Er spins occurring at high temperature. [30]

In this study, we perform a comprehensive study of electronic and thermoelectric transport properties of ErMn$_6$Sn$_6$. Our findings reveal that ErMn$_6$Sn$_6$ exhibits both the topological and anomalous Nernst effects, similar to the electric Hall effects, with a Nernst coefficient of 1.71 µV K$^{-1}$ at 300 K which is on par with other recently reported RMn$_6$Sn$_6$ (R = rare-earth, Y, Lu, Sc) compounds. These results position ErMn$_6$Sn$_6$ as a strong candidate for the development of advanced Nernst effect-based thermoelectric devices, highlighting the significance of exploring magnetic topological materials with interplay among lattice geometry, magnetism and electronic properties.

## 2. **Results and Discussion**

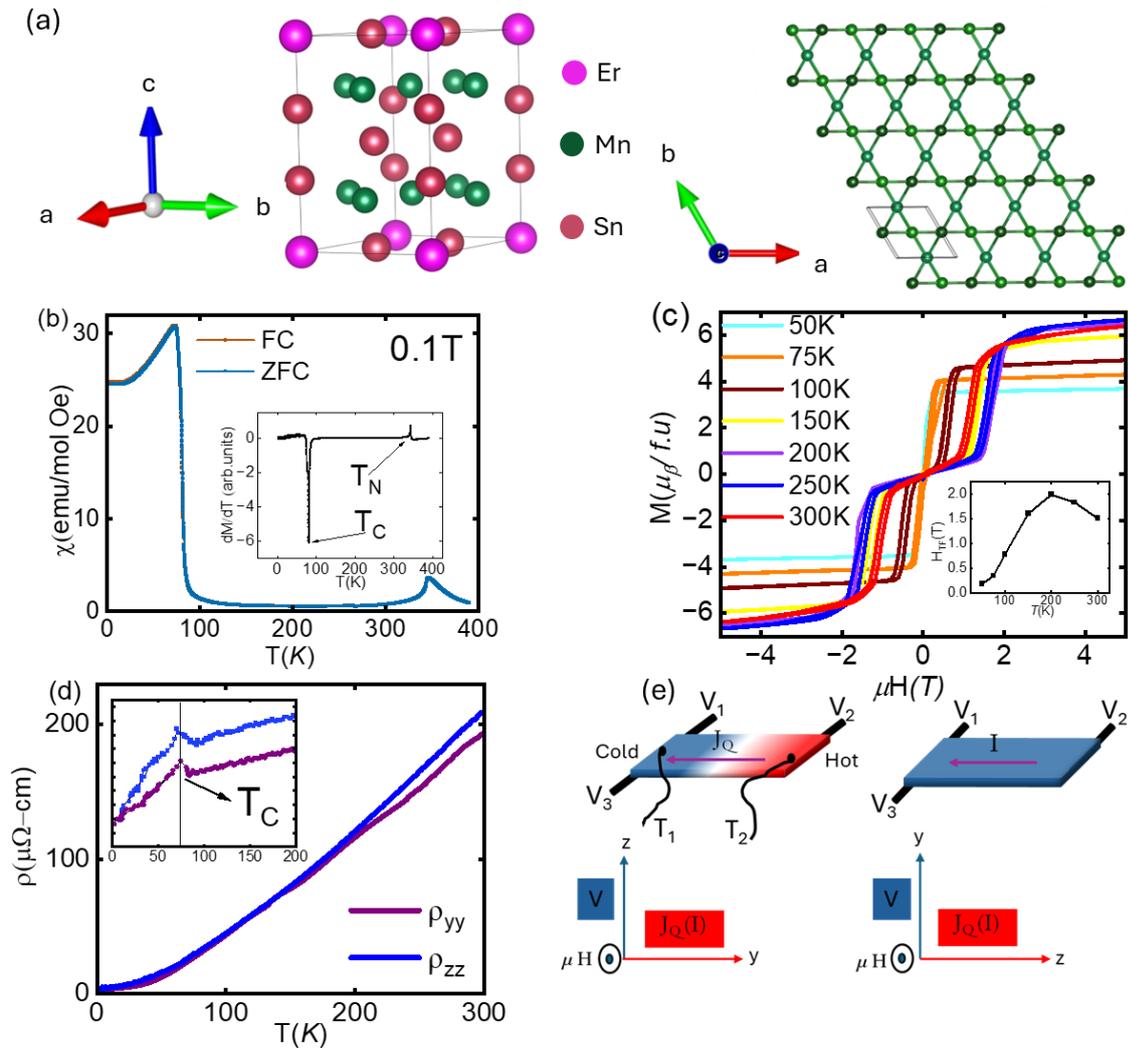

**Figure 1** (a) Top left panel shows the crystal structure of $ErMn_6Sn_6$; top right panel shows the Mn Kagome lattice. (b) Temperature dependence of magnetic susceptibility and its derivative (inset). (c) Magnetic field dependence of magnetization measured at various temperatures; inset shows the temperature dependence of transition field for AFM-to-FIM phase transition. (d) Temperature dependence resistivity and the derivative (inset) measured at zero field. (e)

Schematic setups for electronic and thermoelectric transport measurements. For all magnetic measurements, the magnetic field was applied along the in-plane direction (x-axis).

**Figure** 1(a) depicts the schematic crystal structure of $ErMn_6Sn_6$ visualized using VESTA software. [31] The 1(a) (0, 0, 0) position is occupied by Er atoms, while Mn atoms occupy the 6(i) (1/2, 0, 0.2492) position. Tin (Sn) atoms reside at three distinct Wyckoff sites: $Sn_1$ (1/3, 2/3, 0), $Sn_2$ (1/3, 2/3, 1/2), and $Sn_3$ (0, 0, ~0.8364). [32] Within each unit cell, there are two layers of Mn atoms at z positions of ~ 0.25*$c$ and ~ 0.75*$c$, each of which forms a Kagome web as illustrated in the top right panel. As mentioned previously, due to the interplay among Er-Mn, Er-Er, and Mn-Mn exchange interactions, $ErMn_6Sn_6$ exhibits complex magnetic properties. [33] It undergoes an antiferromagnetic phase transition at $T_N$ = 348 K below which both Mn and Er spins form incommensurate structures along the $c$-axis while ferromagnetically aligned within the $ab$ plane. This is followed by a ferrimagnetic phase transition upon cooling to ~ 68 K, where all Mn spins and Er spins remain ferromagnetically aligned within the $ab$ plane while Mn and Er spins are antiparallel aligned. [29,30,32] In addition, the presence of massive Dirac bands resulting from the Kagome lattice of Mn atoms is found to be close to the Fermi energy, which is anticipated to lead to emergent electronic transport phenomena in $ErMn_6Sn_6$. [29] **Figure** 1(b) presents the temperature-dependent magnetization ($M$) of $ErMn_6Sn_6$ measured with an in-plane magnetic field ($H \parallel ab$) of 0.1 kOe from 2 K to 390 K under zero-field-cooled (ZFC) and field-cooled (FCC) conditions. As shown in the inset, an anomaly observed at $T_N$ ≈ 348 K signifies an antiferromagnetic transition, which is followed by another phase transition at $T_c$ = 68 K with a sharp increase in magnetic moment that corresponds to the ferrimagnetic phase transition, consistent with the literature. [29]

**Figure** 1(c) depicts isothermal magnetization $M(H)$ curves measured at various temperatures. Above $T_c$, a field-induced metamagnetic transition is observed below 2 T, which is accompanied by a large enhancement in magnetization and a hysteresis loop. Above the transition field, the magnetic moment reaches a plateau value of ~ 5-6 $\mu_B$/f.u. only, suggesting a field-induced phase transition from an AFM to a FIM state. [33] This implies a strong Er-Mn antiferromagnetic coupling compared to the Mn-Mn coupling between adjacent planes along the *c*-axis. Interestingly, the transition field shows non-monotonic dependence of temperature, as shown in the inset of **Figure** 1(c), which is consistent with an earlier report. [34] Such a metamagnetic phase transition disappears below $T_c$. It is worth noting that the saturated magnetization decreases upon decreasing the measurement temperature. This is due to the continuously enhanced Er magnetic moment at lower temperature which tends to antiparallel align to the Mn moment due to the Er-Mn antiferromagnetic coupling. [34]

**Figure** 1(d) shows the temperature dependence of electrical resistivity ($\rho_{yy}$ and $\rho_{zz}$) measured at zero magnetic field with the current applied along both y (*I* ∥ y) and z (*I* ∥ z) directions. For both current directions, the resistivity increases monotonically with temperature. The residual resistivity ratio (RRR), defined as $R_{300K}/R_{2K}$, is found to be 144.66 for (*I* ∥ y) and 68.37 for (*I* ∥ z), affirming high quality of $ErMn_6Sn_6$ crystals. There are a couple of features worth pointing out. First, as shown in the inset which displays the temperature derivative of resistivity (*dρ/dT*), a distinct change in the slope of ρ(*T*) is observed at $T_c$, which coincides with the ferrimagnetic transition and implies the magnetoelectric coupling. Second, $\rho_{yy}$ and $\rho_{zz}$ are comparable in magnitude, suggesting three-dimensional nature of the electronic transport properties.

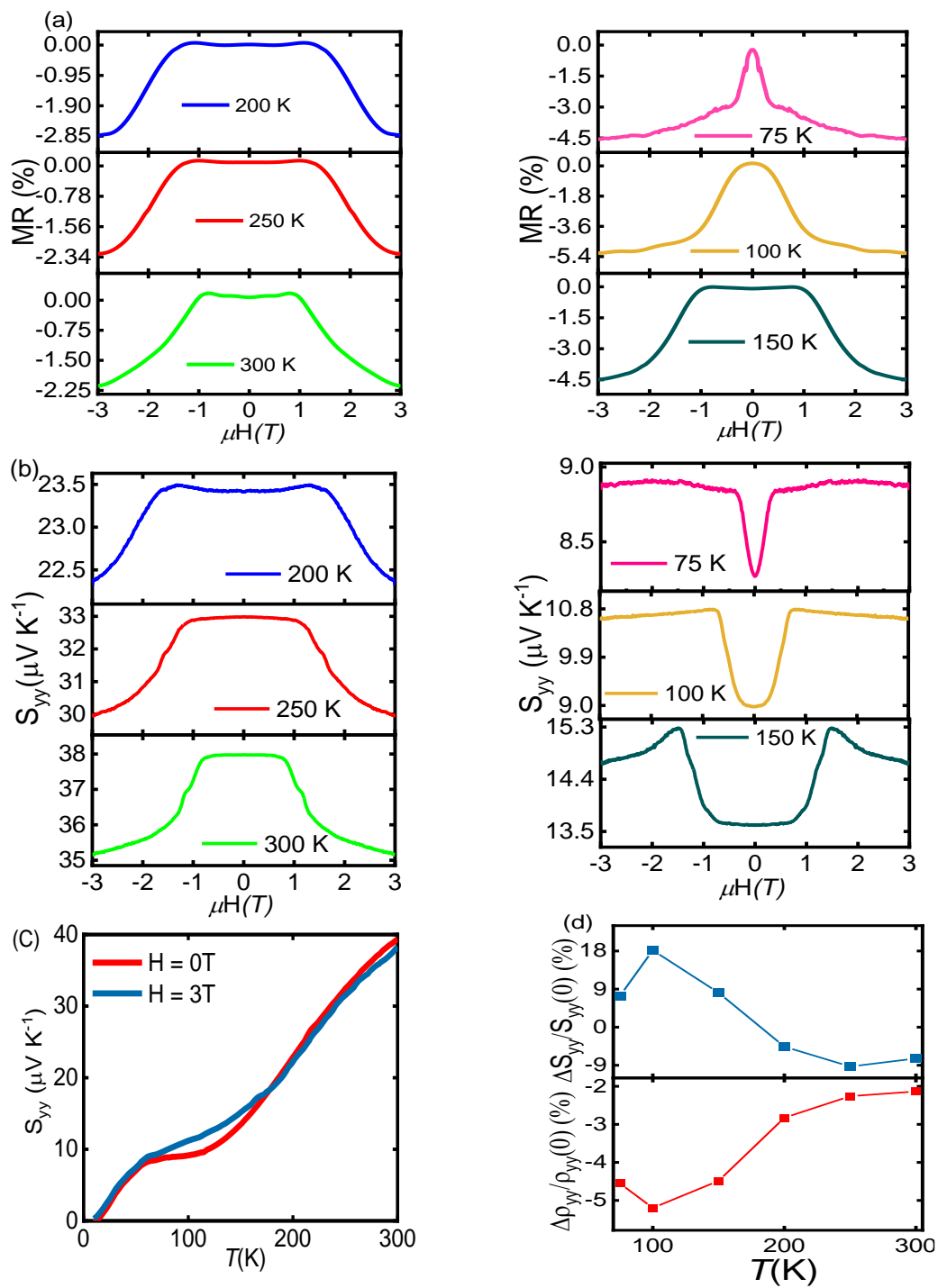

**Figure 2**. Magnetic field dependence of magnetoresistance (MR) (a) and magneto-Seebeck (b) effects measured at various temperatures. (c) Temperature dependence of Seebeck measured

at 0 T and 3 T. (d) Temperature dependence of MR and Seebeck measured at 3 T. For all magnetic measurements, the magnetic field was applied along the in-plane direction (x-axis).

We next present the magnetic field dependence of longitudinal electric and thermoelectric responses of ErMn$_6$Sn$_6$. **Figure** 2(a) shows the field dependence of longitudinal magnetoresistance (MR) which is defined as MR = ($\rho$(H) – $\rho$(0))/$\rho$(0), where $\rho$(H) and $\rho$(0) are the resistivity measured with and without magnetic field, respectively. Both the magnetic field and electric current were applied along the y-axis. At $T_c < T < T_N$, negative MR with respect to the field is observed with the magnitude of MR increasing sharply upon the spin-flop transition as observed in isothermal M(H) measurements shown in **Figure** 1(c). Such a feature is ascribable to the decrease of spin scattering due to the field-induced transition from an incommensurate AFM phase to FIM phase. In contrast, a large positive MR but not field-induced phase transition is observed below $T_c$, as shown in **Figure** S1, [35] which may be related to the dominant orbital scattering. [36] Note that similar MR behaviours above and below $T_c$ has been observed in Er$_{0.5}$Ho$_{0.5}$Mn$_6$Sn$_6$. [37] **Figure** 2(b) presents the magnetic field dependence of Seebeck coefficient ($S_{yy}$) measured at various temperatures. In the temperature range 200 K < $T < T_N$, the behaviour of $S_{yy}$ resembles that of MR with negative magneto-Seebeck effect. That is, the magnitude of $S_{yy}$ decreases at high field and it is smaller than the value at zero field. However, a key distinction emerges in the $T_c < T < 200$ K region where $S_{yy}$ exhibits an upturn and positive magneto-Seebeck feature that is absent in the MR data.

It is known that the thermoelectric response in metallic systems can be approximated using the Mott relation [38]: $S = -\frac{\pi^2 k_B^2 T}{3e\sigma}\frac{d\sigma}{d\zeta}|_{E_F}$, where k$_B$ Boltzmann constant, $\sigma$ is the conductivity, e is electron charge, $\zeta$ represents the energy variable and E$_F$ is the Fermi energy. If $\frac{d\sigma}{d\zeta}|_{E_F}$ remains nearly field-independent, S is then proportional to the resistivity $\rho$ (=1/$\sigma$), resulting in similar magnetic field dependences for the MR and the magneto-Seebeck effect.

This behaviour is indeed observed in ErMn$_6$Sn$_6$ above 200 K, as shown in the left panels of **Figure** 2(a) and (b). However, below 200 K, the MR and magneto-Seebeck effect begin to exhibit an opposite trend, as seen in the right panels of **Figure** 2(a) and (b). A plausible explanation is that $\frac{d\sigma}{d\zeta}|_{E_F}$ becomes field-dependent and adopts an opposite trend for the MR and magneto-Seebeck effect. This may arise from alterations in the electronic structure as ferromagnetic correlations between Mn spins become enhanced upon further cooling. Such a scenario aligns with the temperature dependence of the critical field for the spin-flop phase transition, which peaks around 200 K (inset of **Figure** 1(c)). To fully elucidate the underlying mechanisms driving the observed magneto-Seebeck effect below 200 K (**Figure** 2(b)), further investigations, including detailed calculations or measurements of the electronic structure across different temperature regimes, are highly desirable. Similar magnetic field dependence of longitudinal electric and thermoelectric responses of ErMn$_6$Sn$_6$ was also measured along z direction, as shown in **Figure** S2. [35] **Figure** 2(d) summarizes the temperature dependence of MR and magneto-Seebeck measured at 3 T.

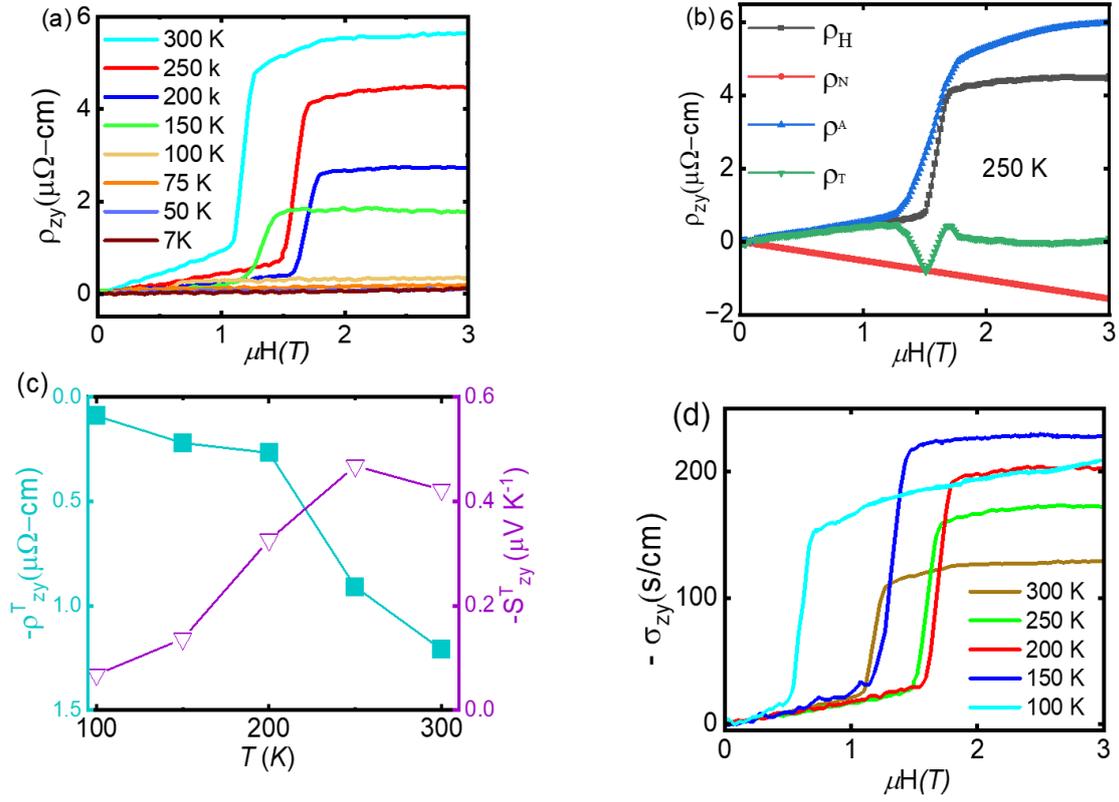

**Figure 3**. Magnetic field dependence of Hall resistivity measured at different temperatures. (b) Magnetic field dependence of different Hall resistivity components at 250 K. (c) Temperature dependence of topological Hall resistivity ($\rho^T_{zy}$) and topological topological Nernst coefficients ($S^T_{zy}$) (d) The magnetic field dependence of Hall conductivity measured at various temperatures. For all magnetic measurements, the magnetic field was applied along the in-plane direction (x-axis).

Now we move on to discuss the transverse electric and thermoelectric response in ErMn$_6$Sn$_6$. In **Figure** 3(a) we present the transverse resistivity response ($\rho_{zy}$) as a function of the magnetic field applied along the x-axis, with the current applied along the y-axis and the transverse voltage measured along the z-axis. The overall behaviour of $\rho_{zy}(H)$ is similar to the $M(H)$ data depicted in **Figure** 1(c). In magnetic topological materials, the total Hall resistivity ($\rho_H$) is expressed as $\rho_H = \rho_N + \rho_A + \rho_T$, where $\rho_N$, $\rho_A$, and $\rho_T$ represent the normal, anomalous, and topological Hall contributions, respectively. The normal Hall contribution $\rho_N(H)$ is defined as $\rho_N(H)=R_0 H$, where $R_0$ is the coefficient of normal Hall resistivity. The sign of $R_0$ is crucial for determining the majority charge carrier. [39] The anomalous Hall resistivity, $\rho_A$, is defined as $\rho_A = R_S 4\pi M$, which can have extrinsic contributions from skew scattering and/or side-jump scattering and/or intrinsic contribution from the Berry curvature of electronic band structure. [39] The topological Hall resistivity $\rho_T$ arises from non-zero spin chirality. [40,41]

In the high-field saturation region where the $\rho_T$ term diminishes, $\rho_H$ simplifies to $\rho_H = R_0 H + R_S 4\pi M$. From the linear plot of $\rho_H/M$ versus $H/M$ in this region, we can estimate the slope $R_0$ and the intercept $4\pi R_S$. As an example, following this method we extracted the Hall contributions of $\rho_N$, $\rho_A$, and $\rho_T$ at 250 K, as shown in **Figure** 3(b). Similar features of Hall effect in ErMn$_6$Sn$_6$ have been reported in. [29] As shown in **Figure S3** [35] and discussed in the Supplemental Information [35], we find that the extracted $\rho_A$ component in ErMn$_6$Sn$_6$ is mainly attributed to the Berry curvature of its electronic structure. **Figure** 3(c) showing the temperature dependence of the maximized $\rho_T$ ($\rho_T^{max}$) extracted at various temperatures. [41] One can see that $\rho_T^{max}$ reaches a maximum at 300 K with a value of 1.2 $\mu\Omega$ cm at 300 K, comparable to the values found in YMn$_6$Sn$_6$ [13,14] and larger than those reported for materials such as Mn$_5$Si$_3$ [42] and Mn$_3$Ga. [43]

The topological Hall effect is generally believed to occur due to the static scalar spin chirality (SSC) associated with non-coplanar magnetic structures. [44–46] Such a mechanism was recently proposed to account for the observed topological Hall effect in isostructural YMn$_6$Sn$_6$ owing to the modification of spiral spin structure in the presence of magnetic field that breaks the spin coplanarity. [13] On the other hand, a fluctuation-driven mechanism was proposed [14], where an internal dynamic Skyrmion-like chiral field is generated through the chiral fluctuations with a field-induced preferential handedness. This consequently leads to non-zero spin chirality, giving rise to a topological Hall effect. In YMn$_6$Sn$_6$ compounds, the topological Hall effect is observed at finite temperatures and within a magnetic phase characterized by a transverse conical spiral (TCS) propagating along the *c*-axis. **Figure** 3(c) shows that the topological Hall effect in ErMn$_6$Sn$_6$ persists in the 100 – 300 K temperature range and a narrow field range between 1 – 1.5 T, a region in which the TCS phase is formed as revealed by recent neutron diffraction studies. [47] While it is challenging to determine which mechanism dominates, it is worth mentioning that the field-induced TCS phase is pivotal for the observation of the topological Hall effect.

**Figure** 3(d) shows the field dependence of the calculated Hall conductivity for different temperatures. The Hall conductivity value is comparable to that of TbMn$_6$Sn$_6$. [10] It is worth noting that similar features and magnitude of electric Hall signal has been measured with the current applied along z-axis and Hall signal measured along the y-axis, i.e, $\rho_{yz}$, which is presented in **Figure** S4, [35] suggesting that the Onsager reciprocal relation is obeyed. [48–50]

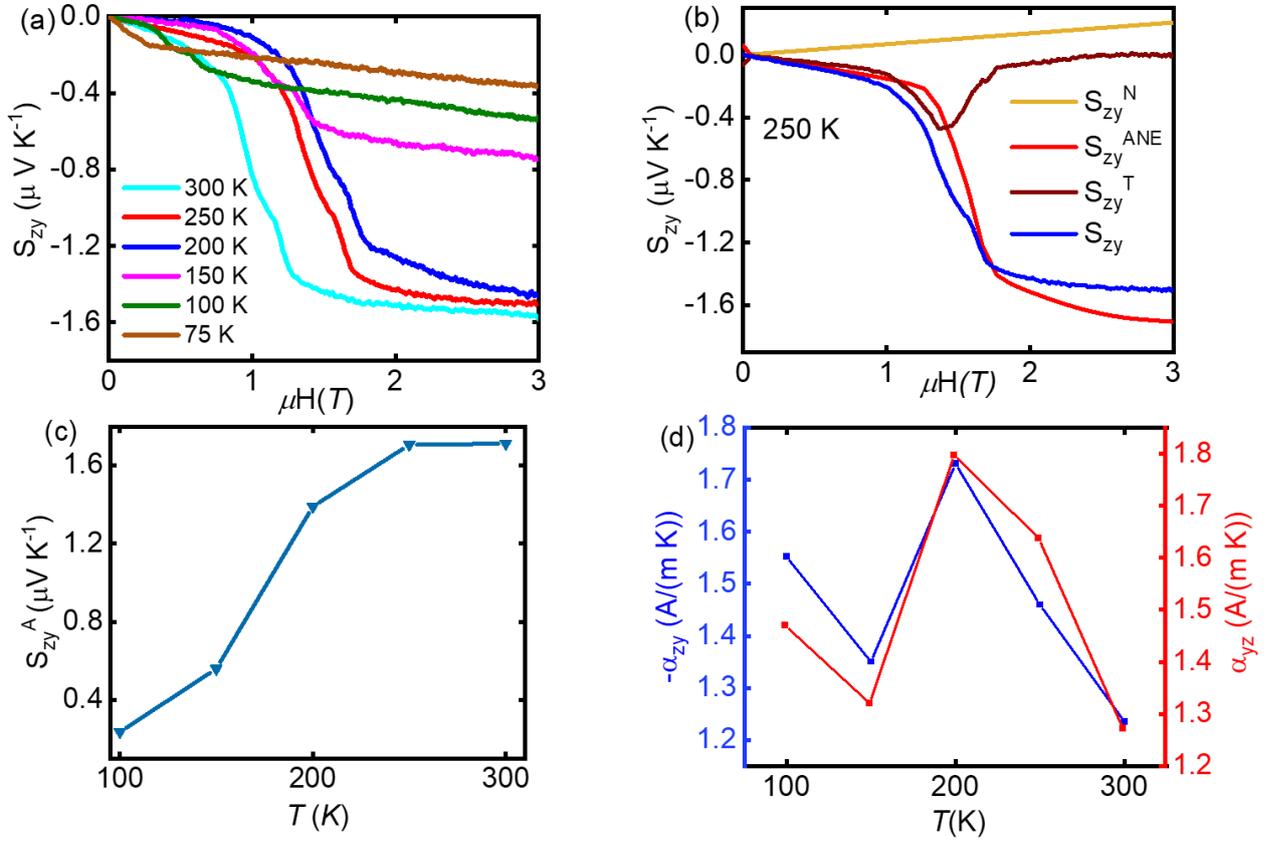

**Figure 4**. (a) Magnetic field dependence of Nernst coefficients measured at various temperatures. (b) Magnetic field dependence of different Nernst effect components at 250 K. (c) Temperature dependence of anomalous Nernst coefficient extracted at 3 T. (d) Temperature dependence of the calculated thermoelectric conductivity at 3T as described in the main text. For all magnetic measurements, the magnetic field was applied along the in-plane direction (x-axis).

Figure 4(a) presents the transverse thermoelectric response of ErMn$_6$Sn$_6$ at various temperatures. The data is plotted as a function of magnetic field applied along the x-axis, with the heat applied along the y-axis and the transverse voltage measured along the z-axis (refer to **Figure** S5 [35] for data measured with the heat applied along the z-axis). The field dependent Nernst coefficient ($S_{zy}$) in ErMn$_6$Sn$_6$, similar to the Hall data presented in **Figure** 3(a), mimics the $M(H)$ data shown in **Figure** 1(c). This similarity underscores the shared underlying mechanisms of the transverse transport properties observed in both Hall and Nernst effects. As to be discussed, the origin of these transverse transport properties stems from the combination of both non-trivial band topology and finite spin chirality of the material. Similar to the electric Hall effect, the Nernst signal (Szy) contains three components $S_{zy} = S^0_{zy} + S^A_{zy} + S^T_{zy}$, where $S^0_{zy}$ represents the normal component, $S^A_{zy}$ is the anomalous contribution, and $S^T_{zy}$ denotes the topological contribution. Using the methodology discussed above, we can extract these three Nernst components as illustrated in **Figure** 4(b) for the data measured at 250 K and temperature dependence of $S^T_{zy}$ is presented in **Figure** 3 (c). The temperature dependence of $S^A_{zy}$ is shown in **Figure** 4(c). The $S^A_{zy}$ increases with temperature, reaching a maximum value of 1.71 µV K$^{-1}$ at 300 K. This value is comparable to other members of the same series, such as 2.2 µV K$^{-1}$ for TbMn$_6$Sn$_6$, [10] 2 µV K$^{-1}$ for YMn$_6$Sn$_6$, [27] and 2.21 µV K$^{-1}$ for ScMn$_6$Sn$_6$ [28] and it surpasses that of previously reported state-of-the-art canted AFM materials, including Mn$_3$Sn [24] and Mn$_3$Ge. [51]

As discussed previously, in this system the massive Dirac bands associated with the Kagome lattice of Mn atoms give rise to non-zero Berry curvature. As a result, charge carriers acquire an 'anomalous velocity' leading to the observed anomalous transverse conductivities. Although both the AHE and the ANE stem from the non-trivial band topology, the ANE is more sensitive than the AHE to the Berry curvature near the Fermi level, making it a more sensitive probe of the electronic structure near Fermi level. On the other hand, both the topological Nernst effect and the topological Hall effect observed in ErMn$_6$Sn$_6$ arise from finite spin chirality, which produces superimposed geometrical effects above the normal Hall and Nernst components. Similar effects have been observed in ScMn$_6$Sn$_6$,[28] although the topological Nernst effect is surprisingly absent in YMn$_6$Sn$_6$ [27]. This discrepancy could be due to subtle differences in the spin structure within the spiral ordering, necessitating careful analysis.

To further understand the thermoelectric properties, we estimated the Nernst thermoelectric conductivity ($\alpha_{ij}$) based on the measured resistivities ($\rho_{ii}$, $\rho_{ij}$) and thermopowers ($S_{jj}$, $S_{ij}$), using the relation: $\alpha_{ij} = S_{ij}\sigma_{ii}+S_{jj}\sigma_{ij}$ [48]. The values of $\alpha_{ij}$ for both yz and zy measurement configurations were found to increase with decreasing temperature from 300 K, as shown in **Figure** 4(d), with a maximum observed at 200 K. This behaviour is consistent with the observations for YMn$_6$Sn$_6$ [27], while an opposite trend is noted for ScMn$_6$Sn$_6$ [28]. It is also noteworthy that $|\alpha_{yz}| = |\alpha_{zy}|$, confirming that the Onsager reciprocal relation [48–50] holds well for ErMn$_6$Sn$_6$. The $\alpha_{yz}$ value peaks at 1.79 A/(m K) at 200 K, surpassing conventional ferromagnets [52–54] and aligning with other topological materials. [20,21,27,28] The Nernst effect enables a more straightforward thermoelectric module design compared to the Seebeck effect, with its orthogonal voltage output. Our study reveals the pronounced Nernst effect in ErMn$_6$Sn$_6$, establishing it as a leading contender for next-generation thermoelectric devices.

Finally, we would like to address the distinction between ErMn$_6$Sn$_6$ and the other RMn$_6$Sn$_6$ compounds. When compared to other RMn$_6$Sn$_6$ compounds, such as YMn$_6$Sn$_6$ [5,13,14,27] and TbMn$_6$Sn$_6$ [9–12], ErMn$_6$Sn$_6$ stands out due to its unique magnetic properties. While YMn$_6$Sn$_6$ and TbMn$_6$Sn$_6$ exhibit an AFM phase and ferrimagnetic phase over the whole temperature regime, respectively, ErMn$_6$Sn$_6$ undergoes a transition from an incommensurate AFM phase to a collinear FIM phase upon cooling. These transitions are driven by the distinctive exchange interactions between Er-Mn and Mn-Mn ions, which influence both the spin configurations and the electronic structure of the material that in turn modulates the transport properties as described below.

For longitudinal transport properties, as discussed previously, above 200 K, magnetoresistance (MR) and magneto-Seebeck are correlated, driven by the suppression of spin scattering in the incommensurate AFM phase. However, below 200 K, as the system gradually approaches the ferrimagnetic phase, the combination of enhanced ferromagnetic correlations, field-induced changes in the electronic structure, and reduced thermal fluctuations modifies the scattering mechanisms, which consequently lead to a sign reversal in magneto-Seebeck and a subtle slope change in MR. Such features are not observed in other $RMn_6Sn_6$ compounds. For the transverse transport properties, the field-induced TCS phase in $ErMn_6Sn_6$ results in a finite spin chirality. This gives rise to a sizable topological Hall effect and topological Nernst effect within this regime. This region represents the boundary between the incommensurate AFM phase — where $ErMn_6Sn_6$ exhibits similarities to $YMn_6Sn_6$ — and the FIM phase — where it aligns more closely with $TbMn_6Sn_6$. A key feature of $ErMn_6Sn_6$ is the observation of both topological Hall effect and topological Nernst effect in the low-field regime, with peaks occurring at ~1.5 T, in contrast to $YMn_6Sn_6$ [13,14] where the topological Hall effect is observed at much higher fields (~5 T) and to $TbMn_6Sn_6$ [10,11] where the topological Hall effect is absent. Above the spin-flop transition, the AHE and ANE are observed in the ferrimagnetic phase in $ErMn_6Sn_6$.

**Figure** S6 [35] summarizes the temperature evolution of longitudinal and transport properties from which an anomaly is observed around 200 K. These longitudinal and transverse transport features underscore the unique coupling between magnetic order (FIM and incommensurate AFM) and electronic topology in $ErMn_6Sn_6$, which not only differentiates $ErMn_6Sn_6$ from other $RMn_6Sn_6$ compounds but also highlights its role as a bridge between the physics observed in the incommensurate AFM compounds (such as $YMn_6Sn_6$) and strongly FIM counterparts (i.e., $TbMn_6Sn_6$).

## 3. Conclusion

Our comprehensive investigation into the electronic and transverse thermoelectric properties of ErMn$_6$Sn$_6$ advances understanding and optimizing magnetic topological materials for thermoelectric applications. The anomalous Nernst effect (ANE), which is highly sensitive to Berry curvature near the Fermi level, demonstrates a notable potential for ErMn$_6$Sn$_6$ with the Nernst coefficient increasing with temperature and reaching a maximum value of 1.71 µV K$^{-1}$ at 300 K. Moreover, ErMn$_6$Sn$_6$ exhibits substantial topological Hall and Nernst effects, which are associated with the non-zero spin chirality existing in the field-induced TCS phase. These features position ErMn$_6$Sn$_6$ as a promising candidate for next-generation thermoelectric devices.

## 4. Methods

Single crystals of ErMn$_6$Sn$_6$ were synthesized using the Sn self-flux method. High-purity Erbium (Er), Manganese (Mn), and Tin (Sn) with a molar ratio of 1:6:20, respectively, were loaded into an alumina crucible which was then sealed in a quartz tube under vacuum [29,55]. ErMn$_6$Sn$_6$ crystallizes in the hexagonal crystal system with a space group of P6/mmm (No. 191) with lattice constants $a = b = 5.527$ Å, $c = 9.020$ Å and crystalline angles $α=β=90°$, $γ=120°$ [32]. The elemental composition of the crystals was confirmed using scanning electron microscopy equipped with energy-dispersive X-ray spectroscopy on a JEOL 7500F ultra-high-resolution instrument. Magnetic susceptibility measurements were performed using a VSM-Superconducting Quantum Interference Device (SQUID) magnetometer. Resistivity and Hall effect measurements were conducted using a Physical Property Measurement System (PPMS). A custom-designed sample puck compatible with the PPMS cryostat was employed for thermoelectric transport measurements. Temperature readings were acquired using type-E (Chromel-Constantan) thermocouples. A Keithley K2182A Nanovoltmeter was utilized to measure the thermoelectric voltage. The cold end of the sample was affixed to a high-conductivity, oxygen-free copper block using silver epoxy. A resistive heater (~1 kΩ) was attached to the opposite (hot) end of the sample, and a heat current (J_Q) was applied within the *yz (xz)*-plane. The external magnetic field was oriented along the *x*-axis (in-plane direction). Note that the *c*-axis is along the *z*-direction, *a*-axis along the *x*-direction, and *b*-axis being orthogonal to *a*-axis in the *ab* plane.

**Acknowledgements**

Olajumoke Oluwatobiloba Emmanuel and Shuvankar Gupta contributed equally to this work. The authors acknowledge the financial support by the U.S. Department of Energy, Office of Science, Office of Basic Energy Sciences, Materials Sciences and Engineering Division under Grant No. DE-SC0019259. The electronic and thermoelectric transport measurements were supported by the National Science Foundation (DMR-2219046)

**Data Availability Statement**

Data for the figures presented in this paper are available at https://doi.org/10.5281/zenodo.15133123

**Supporting Information**

Supporting Information is available from the Wiley Online Library or from the author.

This work investigates the Nernst effect in the Kagome magnet ErMn$_6$Sn$_6$, which exhibits both topological and anomalous Nernst effects with the anomalous Nernst coefficient reaching 1.71 µV K$^{-1}$ at 300 K. This value surpasses that of most canted antiferromagnetic materials, making ErMn$_6$Sn$_6$ a promising candidate for advancing thermoelectric devices based on the Nernst effect.

**ErMn$_6$Sn$_6$: A Promising Kagome Antiferromagnetic Candidate for Room-Temperature Nernst Effect-based thermoelectrics**

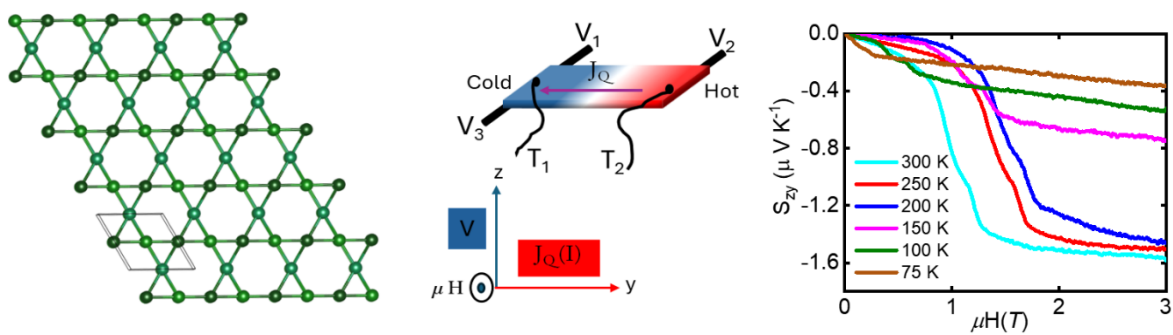

Supporting Information

**Supplemental Information of " ErMn6Sn6: A Promising Kagome Antiferromagnetic Candidate for Room-Temperature Nernst Effect-based thermoelectrics "**

Olajumoke Oluwatobiloba Emmanuel, Shuvankar Gupta* and Xianglin Ke*

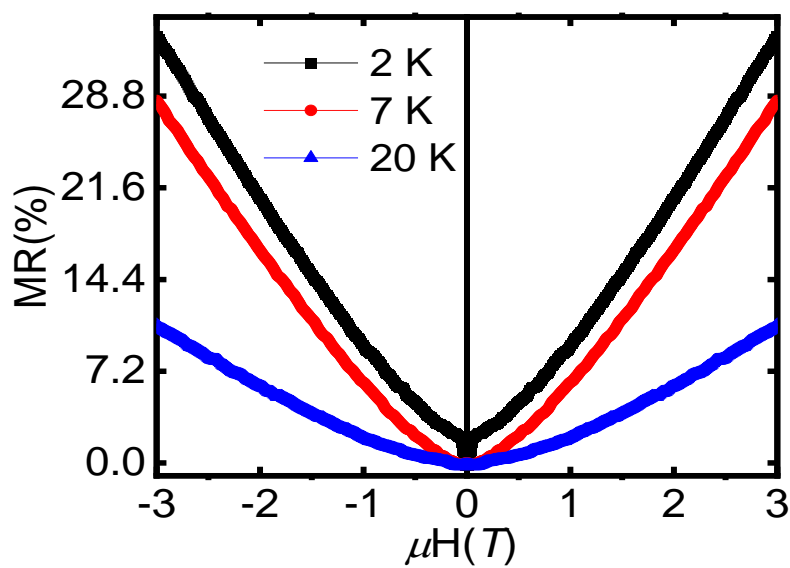

**Figure** S1. Magnetic field dependence of magnetoresistance (MR) measured at several temperatures below $T_c$, magnetic field applied along the in-plane direction.

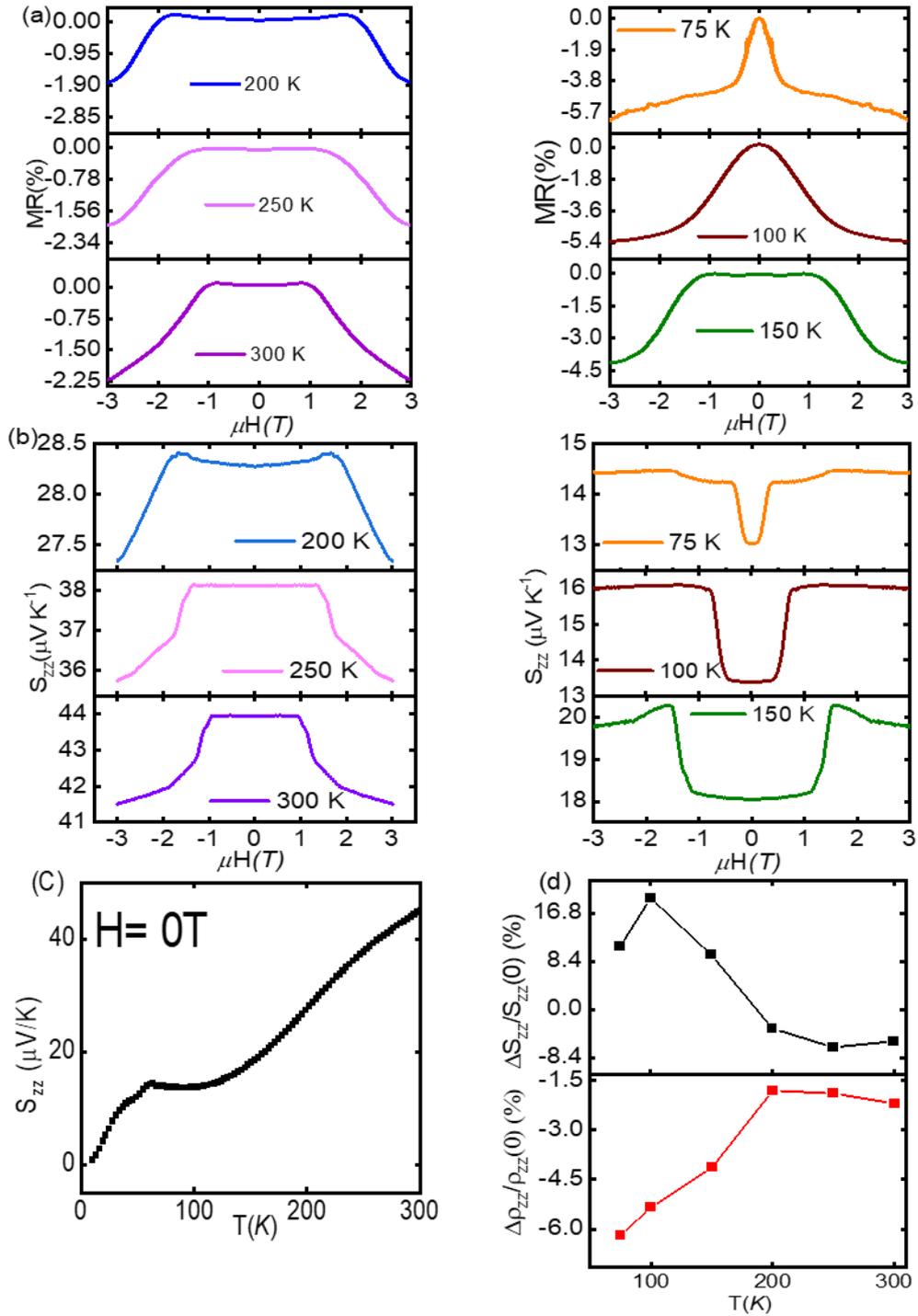

**Figure** S2: Magnetic field dependence along the in-plane direction: of magnetoresistance (MR) (a) and magneto-Seebeck (b) effects measured at various temperatures. (c) Temperature dependence of Seebeck measured at 0 T. (d) Temperature dependence of MR and Seebeck measured at 3 T.

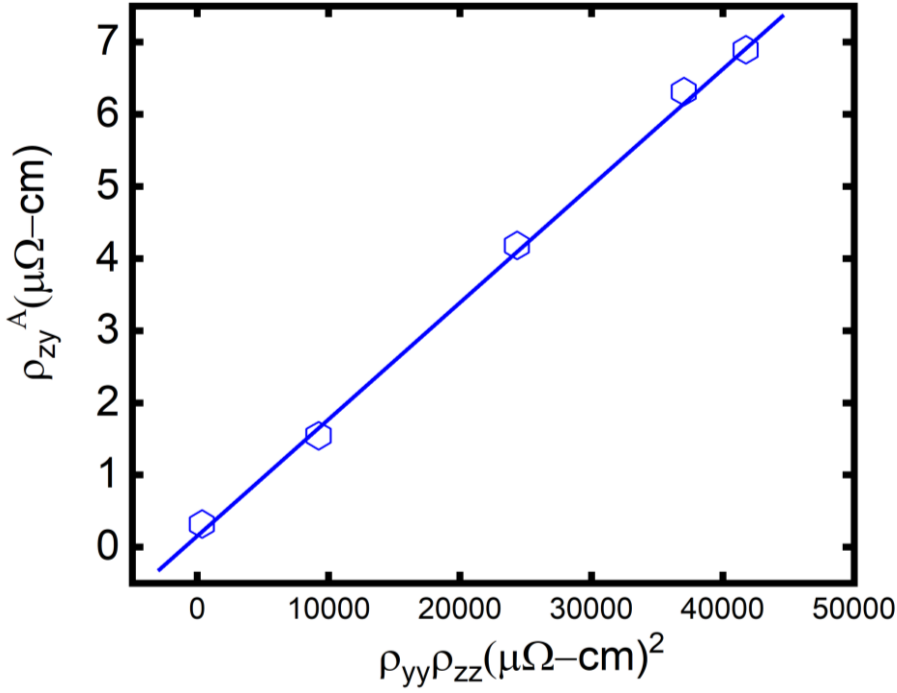

**Figure** S3: Plot of $\rho^A_{zy}$ vs $\rho_{zz}\rho_{yy}$. The blue solid line is the linear fitting.

We analyze the scaling behaviour of $\rho^A_{zy}$ against $\rho_{zz}\rho_{yy}$ the strong linear fit suggests that the extrinsic skew scattering mechanism, which follows the relationship $\rho^A_{zy}$ vs $(\rho_{zz}\rho_{yy})^{1/2}$ can be excluded. Instead, the dominant contributions in ErMn$_6$Sn$_6$ are attributed to either the intrinsic mechanism i.e AHE from Berry Curvature or extrinsic side jump. The extrinsic side-jump contribution of $\rho^A_{zy}$ has been shown to be on the order of $(e^2/(ha))(\varepsilon_{so}/E_F)$, where $e$ is electronic charge, $h$ is the Plank constant, $a$ is the lattice parameter, $\varepsilon_{so}$ is the Spin-orbit coupling (SOC), the $E_F$ is the Fermi energy. For the metallic ferromagnets, the $\varepsilon_{so}/E_F$ is generally smaller than $10^{-2}$ [1] now taking the lattice parameter $a$ = 5.527 Å, the estimated extrinsic side-jump contribution is very small compared to the total AHE $\rho^A_{zy}$, suggest the AHE in ErMn$_6$Sn$_6$ are dominated by intrinsic mechanism.

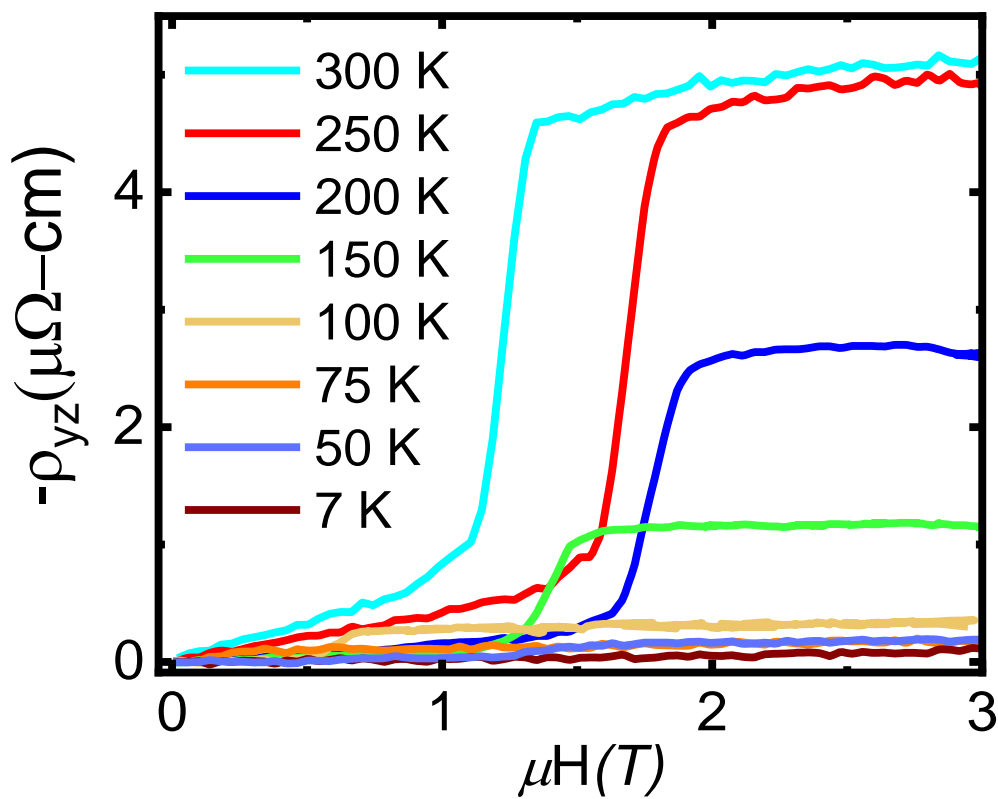

**Figure** S4: Magnetic field dependence of Hall resistivity $\rho_{yz}$ measured at various temperatures, magnetic field applied along the in-plane direction.

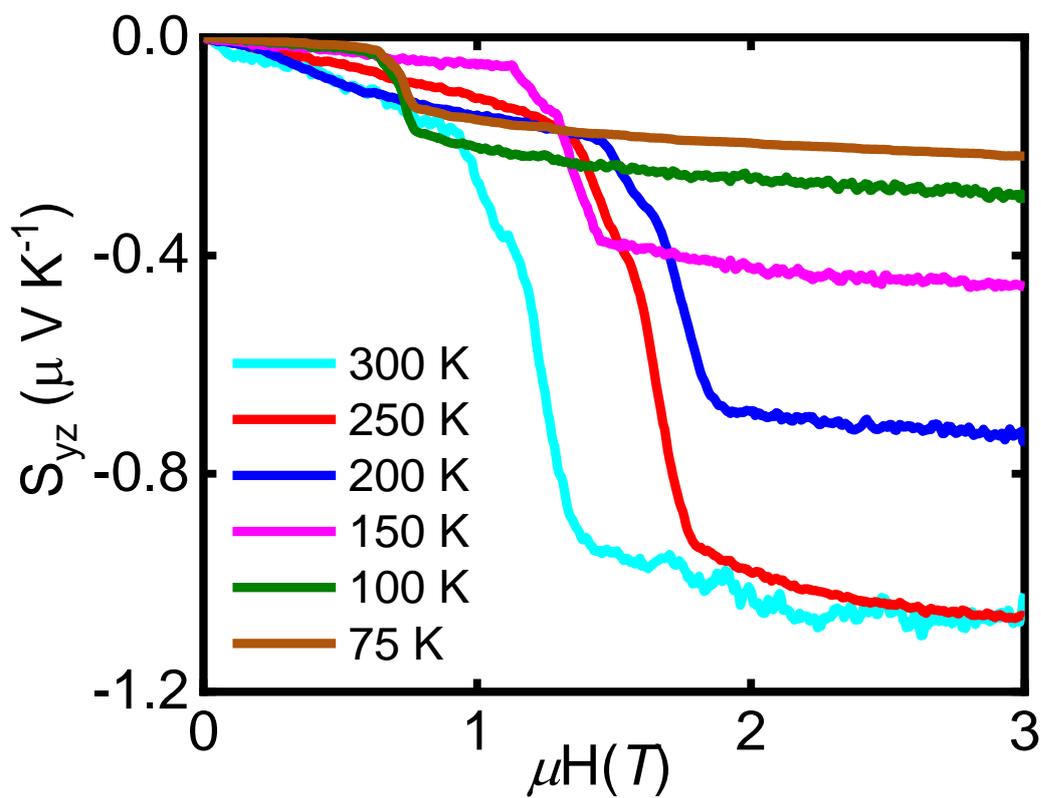

**Figure** S5: Magnetic field dependence of Nernst coefficient $S_{yz}$ measured at different temperatures, magnetic field applied along the in-plane direction.

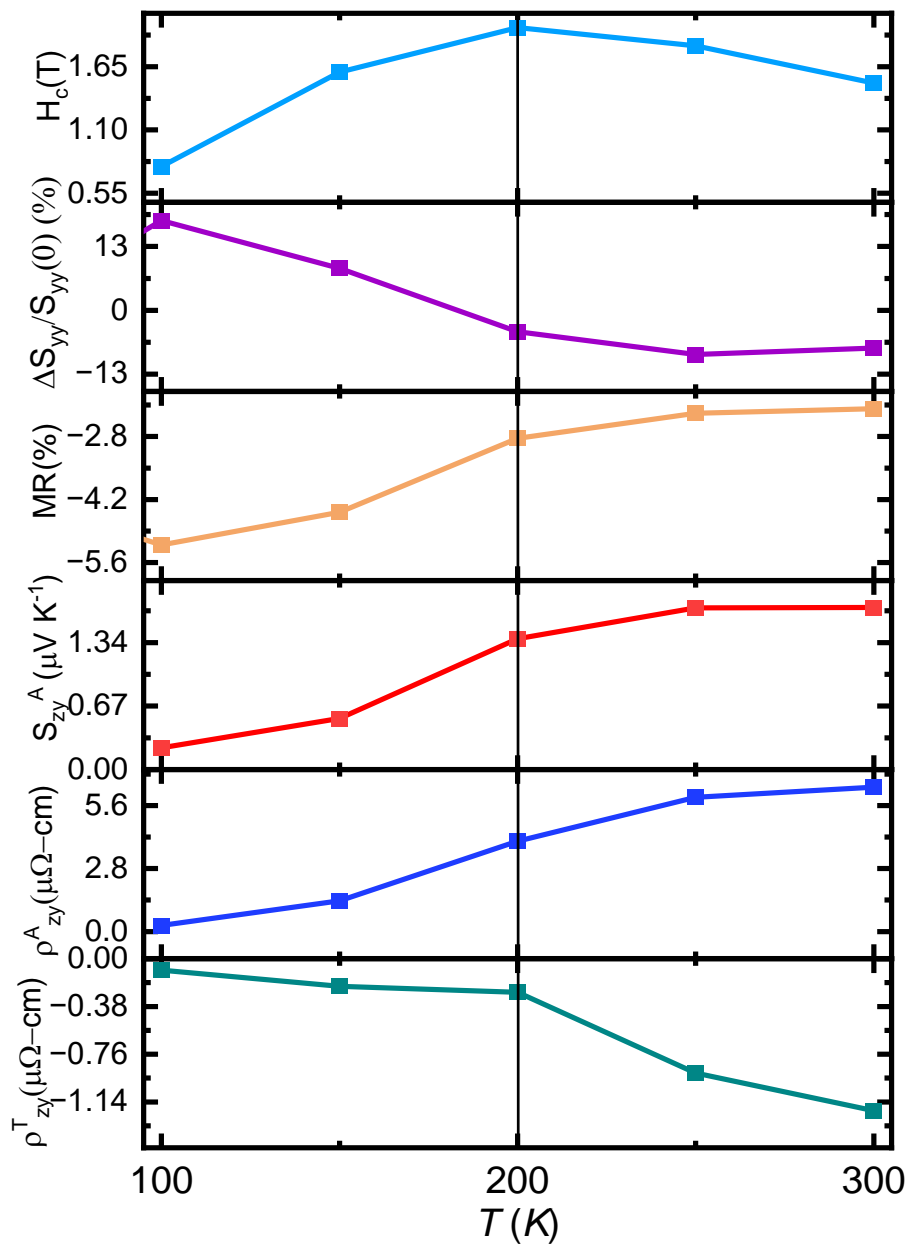

**Figure** S6: Temperature dependence of critical field ($H_c$) of spin-flop transition, magneto-Seebeck (MS), magnetoresistance (MR), anomalous Nernst coefficient ($S^A_{zy}$), anomalous Hall resistivity ($\rho^A_{zy}$), and topological Hall resistivity ($\rho^T_{zy}$). The vertical line serves a guide highlighting an anomaly occurring near 200 K